\documentclass[%
 reprint,
 superscriptaddress,
 amsmath,amssymb,
 aps,
prb,
]{revtex4-2}

\usepackage{graphicx}
\usepackage{dcolumn}
\usepackage{bm}
\usepackage[utf8]{inputenc}
\usepackage[T1]{fontenc}
\usepackage{mathptmx}
\usepackage{etoolbox}
\usepackage{multirow}
\usepackage{float}

\usepackage{soul}
\usepackage{color}
\usepackage{xcolor}

\AtBeginDocument{%
	\newwrite\bibnotes
	\def\bibnotesext{Bi2Te3Bi2Se3.bib}
	\immediate\openout\bibnotes=\jobname\bibnotesext
	\immediate\write\bibnotes{@CONTROL{REVTEX42Control}}
	\immediate\write\bibnotes{@CONTROL{%
			apsrev42Control,author="08",editor="1",pages="1",title="0",year="1"}}
	\if@filesw
	\immediate\write\@auxout{\string\citation{apsrev42Control}}%
	\fi
}%
\begin{document}

\preprint{AIP/123-QED}%

	\title{Ultrafast measurements of mode-specific deformation potentials of Bi$_2$Te$_3$ and Bi$_2$Se$_3$}
	
	\author{Yijing Huang}\email{huangyj@illinois.edu}
 \affiliation{Department of Physics, University of Illinois at Urbana-Champaign, Urbana, IL 61801, USA}
	\affiliation{Stanford PULSE Institute, SLAC National Accelerator Laboratory, Menlo Park, California 94025, USA}
	\affiliation{Department of Applied Physics, Stanford University, Stanford, California 94305, USA}
	\affiliation{Stanford Institute for Materials and Energy Sciences, SLAC National
Accelerator Laboratory, Menlo Park, California 94025, USA}

	\author{José D. Querales-Flores}
	\affiliation{Tyndall National Institute, Dyke Parade, Cork, Ireland}
	
	\author{Samuel W. Teitelbaum}
	\affiliation{Stanford PULSE Institute, SLAC National Accelerator Laboratory, Menlo Park, California 94025, USA}
	\affiliation{Department of Physics, Arizona State University, Tempe, Arizona 85281, USA}
	
	\author{Jiang Cao}
	\affiliation{Integrated Systems Laboratory, ETH Zurich, Zürich, Switzerland}
	
	\author{Thomas Henighan}
	\affiliation{Stanford PULSE Institute, SLAC National Accelerator Laboratory, Menlo Park, California 94025, USA}
	\affiliation{Department of Physics, Stanford University, Stanford, California 94305, USA}
	
	\author{Hanzhe Liu}
	\affiliation{Stanford PULSE Institute, SLAC National Accelerator Laboratory, Menlo Park, California 94025, USA}
	\affiliation{Department of Physics, Stanford University, Stanford, California 94305, USA}
	\affiliation{Department of Chemistry, Purdue University, West Lafayette, Indiana 47907, USA}
	
	\author{Mason Jiang}
	\affiliation{Stanford PULSE Institute, SLAC National Accelerator Laboratory, Menlo Park, California 94025, USA}
	\affiliation{Department of Applied Physics, Stanford University, Stanford, California 94305, USA}
	
	\author{Gilberto De la Pe\~na}
\affiliation{Stanford Institute for Materials and Energy Sciences, SLAC National
Accelerator Laboratory, Menlo Park, California 94025, USA}
\affiliation{Stanford PULSE Institute, SLAC National Accelerator
Laboratory, Menlo Park, California 94025, USA}

		\author{Viktor Krapivin}
	\affiliation{Stanford PULSE Institute, SLAC National Accelerator Laboratory, Menlo Park, California 94025, USA}
	\affiliation{Department of Applied Physics, Stanford University, Stanford, California 94305, USA}
	
		\author{Johann Haber}
	\affiliation{Stanford PULSE Institute, SLAC National Accelerator Laboratory, Menlo Park, California 94025, USA}

	\author{Takahiro Sato}
\affiliation{Linac Coherent Light Source, SLAC National Accelerator Laboratory, Menlo Park, California
94025, USA}

\author{Matthieu Chollet}
\affiliation{Linac Coherent Light Source, SLAC National Accelerator Laboratory, Menlo Park, California
94025, USA}

\author{Diling Zhu}
\affiliation{Linac Coherent Light Source, SLAC National Accelerator Laboratory, Menlo Park, California 94025, USA}

\author{Tetsuo Katayama}
\affiliation{Japan Synchrotron Radiation Research Institute, Kouto 1-1-1, Sayo, Hyogo 679-5198, Japan}
\affiliation{RIKEN SPring-8 Center, 1-1-1 Kouto, Sayo, Hyogo 679-5148, Japan}

	\author{Robert Power}
	\affiliation{Department of Physics, University College Cork, College Road, Cork, Ireland}
	
	\author{Meabh Allen}
	\affiliation{Department of Physics, University College Cork, College Road, Cork, Ireland}
	
	\author{Costel R. Rotundu}
	\affiliation{Stanford Institute for Materials and Energy Sciences, SLAC National Accelerator Laboratory, Menlo Park, California 94025, USA}
	
	\author{Trevor P. Bailey}
	\affiliation{Department of Physics, University of Michigan, Ann Arbor, Michigan 48109, USA}
	
	\author{Ctirad Uher}
	\affiliation{Department of Physics, University of Michigan, Ann Arbor, Michigan 48109, USA}
	
	\author{Mariano Trigo}
	\affiliation{Stanford PULSE Institute, SLAC National Accelerator Laboratory, Menlo Park, California 94025, USA}
	\affiliation{Stanford Institute for Materials and Energy Sciences, SLAC National Accelerator Laboratory, Menlo Park, California 94025, USA}
	
	\author{Patrick S. Kirchmann}
	\affiliation{Stanford Institute for Materials and Energy Sciences, SLAC National Accelerator Laboratory, Menlo Park, California 94025, USA}
	
	\author{\'Eamonn D. Murray}
	\affiliation{Department of Physics and Department of Materials, Imperial College London, London SW7 2AZ, United Kingdom}

	
	\author{Zhi-Xun Shen}
	\affiliation{Stanford Institute for Materials and Energy Sciences, SLAC National Accelerator Laboratory, Menlo Park, California 94025, USA}
	\affiliation{Department of Applied Physics, Stanford University, Stanford, California 94305, USA}
	
	\author{Ivana Savić}
	\affiliation{Department of Physics, King's College London, The Strand, London WC2R 2LS, United Kingdom}
	
	\author{Stephen Fahy}
	\affiliation{Department of Physics, University College Cork, College Road, Cork, Ireland}
	\affiliation{Tyndall National Institute, Dyke Parade, Cork, Ireland}
	
	\author{Jonathan A. Sobota}\email{sobota@slac.stanford.edu}
	\affiliation{Stanford Institute for Materials and Energy Sciences, SLAC National Accelerator Laboratory, Menlo Park, California 94025, USA}
	
	\author{David A. Reis}\email{dreis@stanford.edu}
	\affiliation{Stanford PULSE Institute, SLAC National Accelerator Laboratory, Menlo Park, California 94025, USA}
	\affiliation{Stanford Institute for Materials and Energy Sciences, SLAC National Accelerator Laboratory, Menlo Park, California 94025, USA}
	\affiliation{Department of Applied Physics, Stanford University, Stanford, California 94305, USA}
	\affiliation{Department of Photon Science, Stanford University, Stanford, California 94305, USA}
	
	\begin{abstract}
  Quantifying electron-phonon interactions for the surface states of topological materials can provide key insights into surface-state transport, topological superconductivity, and potentially how to manipulate the surface state using a structural degree of freedom. 
  We perform time-resolved x-ray diffraction (XRD) and angle-resolved photoemission (ARPES) measurements on Bi$_2$Te$_3$ and Bi$_2$Se$_3$, following the excitation of coherent A$_{1g}$ optical phonons. 
  We extract and compare the deformation potentials coupling the surface electronic states to local A$_{1g}$-like displacements in these two materials using the experimentally determined atomic displacements from XRD and electron band shifts from ARPES.
  We find the coupling in Bi$_2$Te$_3$ and Bi$_2$Se$_3$ to be similar and in general in agreement with expectations from density functional theory. 
  We establish a methodology that quantifies the mode-specific electron-phonon coupling experimentally, allowing detailed comparison to theory. 
  Our results shed light on fundamental processes in topological insulators involving electron-phonon coupling. 
	\end{abstract}

\maketitle
\section{Introduction}
Electron-phonon interactions are of fundamental importance in condensed matter.  In the case of  topological insulators the scattering properties of their topological surface states (SS) have a direct impact on their application to energy-efficient electronics~\citep{qu2010quantum,analytis2010two,xiu2011manipulating,hamdou2013surface,hoefer2014intrinsic}.Previous studies employed a variety of methods to study the electron-phonon coupling: self-energy analysis in ARPES~\citep{hatch2011stability,kondo2013anomalous,chen2013tunable,Pan2012,heid2017electron}, helium atom scattering~\citep{ruckhofer2020terahertz,tamtogl2017electron,tamtogl2017electron} and transport measurements~\cite{zhu2012electron}. 
The metal-intercalated Bi$_2$Se$_3$ family of topological insulators are also topological superconductor candidates~\citep{fu2010,fu2014,matano2016spin,asaba2017rotational,yonezawa2017thermodynamic,willa2018nanocalorimetric}, where electron-phonon coupling potentially mediates unconventional pairing symmetries \cite{wang2019evidence}. Therefore, quantifying the mode-specific electron-phonon coupling may benefit the search for and understanding of topological superconductors. 
Studying electron-phonon coupling is also beneficial for understanding and designing functional topological materials. 
For example, the SS has been predicted to facilitate certain chemical reactions\citep{Li2019,Xiao2015}, among which some benefit from phonon participation~\citep{Kroes2008}.
Finally, electron-phonon interactions hold implications for optical control of matter~\citep{tokura2006photoinduced,basov2017towards}, for example, by ultrafast manipulation of SS transport~\citep{luo2019ultrafast}, or by exciting certain lattice degrees of freedom to switch topological phases~\citep{sie2019ultrafast,vaswani2020light,aryal2021topological,Luo2021}. 

The deformation potential (DP) quantifies the electron-phonon coupling of a specific electronic band to a specific phonon mode as the change of electronic energy for a given lattice displacement~\cite{BardeenShockely}. Ultrafast spectroscopy has demonstrated the capability to measure this band- and mode-resolved electron-phonon coupling~\citep{Rettig2015,gerber2017femtosecond}. 
Here, we report ultrafast measurements of band- and mode-specific DPs in the prototypical topological insulators Bi$_2$Te$_3$ and Bi$_2$Se$_3$. 
We measure the DPs of the SS coupled to A$_{1g}$ phonons by combining the atomic displacements measured by XRD with the energy shift of the SS measured by time-resolved ARPES.
Our mode- and band-specific measurement of the DPs contributes to a microscopic understanding of SS transport where electron-phonon coupling is an important contribution to the scattering rates. 
The methodology we lay out could also be useful for the investigation of topological switches induced by structural change, since one can monitor the structural change with XRD while observing the topological state with ARPES.

We find that the experimental DP of the A$_{1g}^{(1)}$ phonon mode is 1.4 times larger in Bi$_2$Te$_3$ than in Bi$_2$Se$_3$. Within experimental uncertainties the DPs are similar, and on the order of $2$~meV/pm. 
This qualitatively agrees with the naive expectation for two isostructural and isoelectronic materials and notably is a factor 5 smaller than in the strongly correlated superconductor FeSe~\citep{gerber2017femtosecond}.
Predictions from density functional theory (DFT) match experimental values within a factor of 2.5. 
DFT correctly predicts a larger DP in Bi$_2$Te$_3$ than in Bi$_2$Se$_3$. Constrained DFT calculations of the excited state reproduce the experimentally observed mode softening well. 
The consistent experimental values and their general agreement with theory validate our experimental and theoretical concepts while also pointing to their challenges.

In the following, we define the DP with respect to A$_{1g}$ modes and the SS in Bi$_2$Te$_3$ and Bi$_2$Se$_3$ as
    \begin{equation}\label{eq:definition of DP}
        \text{DP}= \frac{\Delta E_{\rm SS}^{A_{1g}}}{u_{\rm SS}^{A_{1g}}},
    \end{equation}
where 
$u_{\rm SS}^{A_{1g}}$ is an atomic displacement near the surface projected onto the bulk A$_{1g}$ mode eigenvector (as shown in Fig~\ref{fig:XRD&ARPES} (a)) and averaged over the spatial extent of the SS. 
This definition allows us to formally define a surface deformation potential relative to a bulk phonon, even though the interatomic forces near the surface may be subtly different from true bulk-like forces.
$\Delta E_{\rm SS}^{A_{1g}}$ is the energy shift of the SS induced by the atomic displacement along the A$_{1g}$ mode eigenvector.
Below we choose to quantify $u_{\rm SS}^{A_{1g}}$ with respect to the bismuth displacements for convenience in comparing the two materials. We suppress $A_{1g}$ to simplify the notation in the following.

When combining XRD and ARPES data to calculate DPs, we need to account for the fact that the two methods are respectively bulk- and surface-sensitive probes. 
This is important because the finite penetration depth of the pump laser leads to diffusion of excited carriers and in turn to an inhomogeneous phonon displacement inside the probed volume of the sample. This can lead to systematic errors in the analysis that require correction depending on the experimental details. 
For example, the study in Gerber and Yang \emph{et al.} \cite{gerber2017femtosecond} utilized FeSe films with a thickness comparable to the optical penetration depth, thus a quantitative correction due to diffusion is minor. In contrast, we require substantial corrections for the thicker Bi$_2$Te$_3$ and Bi$_2$Se$_3$ samples studied here. 

The paper is organized as follows: first, we present the time-resolved ARPES and XRD experiments in section II. We then detail the DP calculations in section III, where, guided by simulations, we correct for methodological effects. In section IV, we summarize the DFT calculations. In section V we conclude our findings and suggest systematic improvements for future work.

\section{Experimental details}

	\begin{figure*} 
		\includegraphics{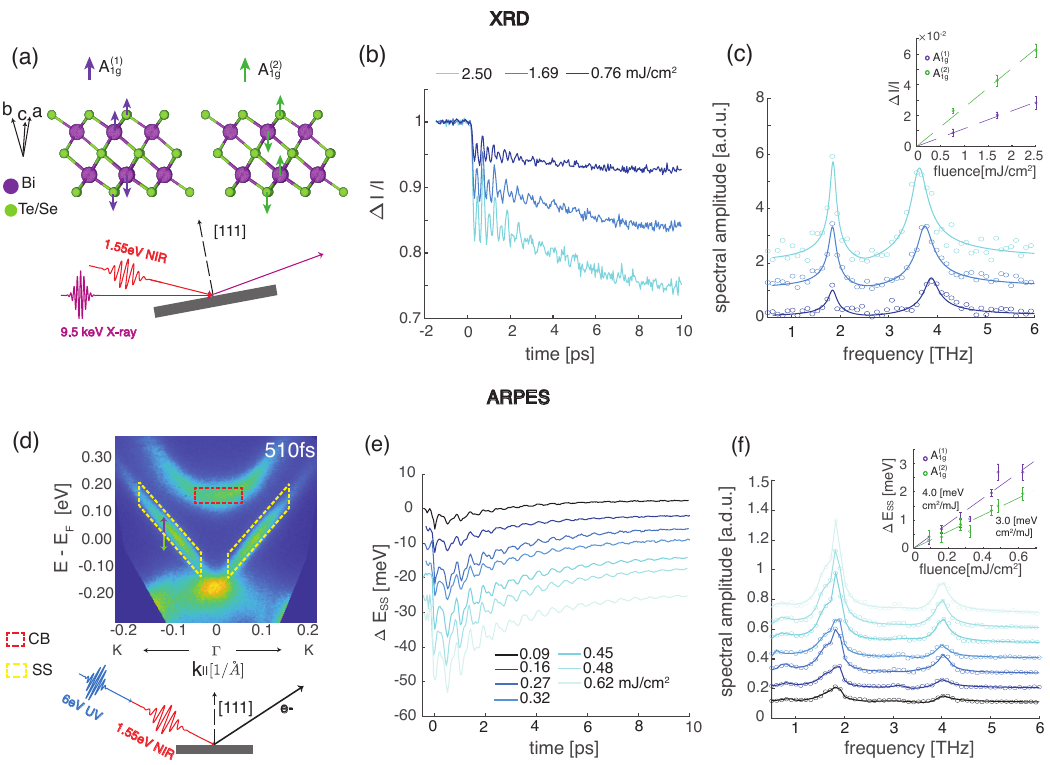}
		\caption{Room temperature time-resolved measurements on Bi$_2$Te$_3$.
		(a-c) Time-resolved XRD. 
		(a) shows two A$_{1g}$ phonon modes of Bi$_2$X$_3$(X=Se,Te), in purple (A$_{1g}^{(1)}$) and green (A$_{1g}^{(2)}$) arrows, illustrating direction of atomic motion (along [111]) upon photoexcitation of the 1.55~eV laser. In the experiment, X-ray and NIR pulses are both grazing incident to match their penetration depth. Sample normal [111] is demonstrated with broken arrow.
		(b) Time-resolved diffraction of Bi$_2$Te$_3$ (556) peak for different fluences. Data taken at LCLS. (c) Fourier transform of (b). Inset shows linear fluence dependence of A$_{1g}^{(1)}$ and A$_{1g}^{(2)}$ mode induced (556) peak diffraction amplitudes $\Delta I/I$. 
		(d-f) Time-resolved ARPES. 
		(d) The electron population at the delay of 510~fs. The boxed area corresponds to a region of interest for the SS (yellow box) and conduction band (CB, red box). 
		In the ARPES experiment, the incident angle of the IR pump and UV probe are both 50 $^{\circ}$. 
		The A$_{1g}$ phonon-induced SS shift directions can be obtained from the initial phase of the coherent oscillation, as indicated by purple(A$_{1g}^{(1)}$) and green(A$_{1g}^{(2)}$) arrows. 
		(e) Averaged energy shift of SS ($\Delta E_{\rm SS}$) in the yellow region of interest in (d). (f) Fourier transform of (e). Inset shows linear fluence dependence of SS energy modulation amplitude $\Delta E_{\rm SS}$ induced by A$_{1g}^{(1)}$ and A$_{1g}^{(2)}$ mode.   
		}
		\label{fig:XRD&ARPES}
	\end{figure*}

In the XRD and ARPES experiments, Bi$_2$Se$_3$ and Bi$_2$Te$_3$ are pumped at room temperature with ultrafast 1.55~eV laser pulses. This excites fully symmetric coherent optical A$_{1g}$ phonons that produce modulations in the XRD intensities from which we extract the amplitudes of phonon motions. The coherent A$_{1g}$ phonons also induce electron band shifts via electron-phonon coupling, producing oscillations in electron band energies which we quantify using time-resolved ARPES. 

We first describe the experiments on Bi$_2$Te$_3$.
The atomic motion diagram in Fig.~\ref{fig:XRD&ARPES} (a) shows the two A$_{1g}$ modes.
To extract the phonon amplitudes, we performed time-resolved XRD at the X-ray Pump-Probe (XPP) instrument of the Linac Coherent Light Source (LCLS) using 9.5~keV photon energy and a 50~nm Bi$_2$Te$_3$ film on sapphire.
Both the optical laser and the X-ray laser are grazing incident on the sample in order to match their penetration depths. 
The time resolution is approximately 50~fs, taking into account the laser and X-ray pulse duration, after the correction of timing jitter between them. 
Fig.~\ref{fig:XRD&ARPES}(b) shows the normalized time-dependent diffracted intensity $\Delta I/I$ of the (556) peak, which is sensitive to both A$_{1g}^{(1)}$ (1.8~THz) and A$_{1g}^{(2)}$ (3.9~THz) modes, as evidenced by the Fourier transform in Fig.~\ref{fig:XRD&ARPES} (c). 
The A$_{1g}$ phonons are launched through the displacive excitation of coherent phonons (DECP), where the quasi-equilibrium phonon coordinates are suddenly shifted upon photoexcitation~\cite{zeiger1992}. The photoexcited forces drive the phonon motion along [111] (see Appendix A for the rhombohedral primitive unit cell), illustrated with the purple (A$_{1g}^{(1)}$) and green (A$_{1g}^{(2)}$) arrows in Fig.~\ref{fig:XRD&ARPES} (a). 
The time-dependent intensity modulations can be described as damped harmonic oscillators, each parameterized by an amplitude, a frequency, a damping constant and an initial phase. We extract these parameters using a linear prediction singular value decomposition (LPSVD) method~\citep{barkhuijsen1985retrieval,led1991application}. The inset of Fig~\ref{fig:XRD&ARPES} (c) shows the fluence dependence of the normalized amplitude $\Delta I/I$.
We constrain the linear fits (broken lines in the inset) to intersect the origin because the coherent atomic motion amplitude is zero without optical excitation. The intensity modulation amplitude is converted to vibrational mode amplitude using the phonon eigenvector experimentally determined by measuring multiple Bragg peaks (see Appendix A).

To extract the electron band energy shifts induced by coherent phonon motion, we performed time-resolved ARPES measurements on Bi$_2$Te$_3$ in our time-resolved ARPES laboratory. A single crystal of Bi$_2$Te$_3$ was cleaved in ultra-high vacuum and probed using 6~eV UV laser with an overall time resolution of 60~fs. The photoexcited ARPES spectrum at a delay time of 510~fs is shown in Fig.~\ref{fig:XRD&ARPES}(d), with the SS (yellow) and conduction band (CB, red) denoted by the boxed regions. We fit the peak position of each energy distribution curve (EDC) as a function of delay and parallel momentum $\textbf{k}_{\vert\vert}$ for the SS along $\Gamma$-K. We find negligible dependence on $\textbf{k}_{\vert\vert}$ (see Appendix B) and therefore average the time-resolved energy shifts over $\textbf{k}_{\vert\vert}$ within the indicated region to maximize the signal-to-noise ratio for the subsequent analysis. We note that the bulk states disperse in three dimensions, and there is substantial uncertainty in the probed out-of-plane momentum component $\textbf{k}_{\bot}$, which precludes a quantitative analysis of the bulk bands~\citep{Xiong2017}.
In contrast, the SS has no $\textbf{k}_{\bot}$-component and is therefore not subject to this ambiguity. 
Fig.~\ref{fig:XRD&ARPES}(e) shows the $\textbf{k}_{\vert\vert}$-averaged energy shifts of the SS as a function of fluence.
The Fourier transform in Fig.~\ref{fig:XRD&ARPES}(f) shows two main A$_{1g}^{(1)}$ and A$_{1g}^{(2)}$ peaks with shoulders associated with the surface termination~\cite{Sobota_2014_distinguishing,surfacestate2023}. 
The main peaks have the same frequency as the modes coupling to the bulk conduction band \cite{surfacestate2023} and we assign them to bulk-like $q=0$ modes which can be compared to the modes measured by XRD.
The amplitudes extracted from LPSVD are linearly dependent on fluence, as shown in the Fig.~\ref{fig:XRD&ARPES} (f) inset. We checked the results of LPSVD by comparing the fitted spectra to the real and imaginary parts of the measured Fourier spectrum and found good agreement.  
    
We now discuss the experimental details for measuring the deformation potential of Bi$_2$Se$_3$. We performed diffraction measurements at SACLA-FEL on both Bi$_2$Se$_3$ and  Bi$_2$Te$_3$. The pump laser is 1.55~eV, which was approximately colinear with the 9.5~keV hard X-ray at grazing incidence to the sample. The time resolution is estimated to be 50~fs. In this experiment, Bi$_2$Se$_3$ was a 90nm film on BaF$_2$ while Bi$_2$Te$_3$ was a 200nm film on sapphire.
The two thin film samples were mounted side by side with their surface normals aligned to within the order of 0.01$^\circ$. This alignment ensures that systematic errors in the incident fluence calibration between the two materials are negligible in the XRD experiment. 

	\begin{figure} 
	\includegraphics{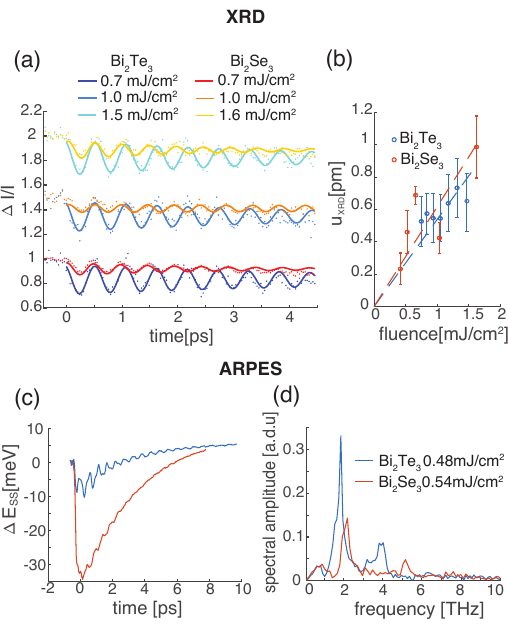}
	\caption{ Room temperature time-resolved measurements comparing photoexcited Bi$_2$Te$_3$ and Bi$_2$Se$_3$. 
	(a) Bi$_2$Te$_3$ and Bi$_2$Se$_3$ time- resolved XRD on peak (445) measured at SACLA. 
	The data are grouped into three levels of similar absorbed fluences. The intensity modulation of Bi$_2$Te$_3$ is consistently around three times larger than Bi$_2$Se$_3$.
	(b) Linear fluence dependence of $u ^{\mathrm{Bi}}$. The similar slopes of linear fits on Bi$_2$Te$_3$ and Bi$_2$Se$_3$ show that their Bi atoms are driven at similar amplitudes. 
	(c) Bi$_2$Te$_3$ and Bi$_2$Se$_3$ time-resolved ARPES measurements of SS at similar absorbed fluences. 
	(d) is the Fourier transform of (c).
	}
	\label{fig:SACLA_amplitude}
	\end{figure}

Fig.~\ref{fig:SACLA_amplitude}(a) shows raw data (dots) and fit (solid lines) of the Bi$_2$Te$_3$ and Bi$_2$Se$_3$ diffraction intensities on Bragg peak (445). Data is grouped by similar absorbed fluence after accounting for differences in the film thickness and optical absorption (see Appendix D), and offset for clarity. Since the A$_{1g}^{(2)}$ mode was not resolved in Bi$_2$Se$_3$, we only present the atomic motion of the A$_{1g}^{(1)}$ mode. 
From Fig.~\ref{fig:SACLA_amplitude}(a), the diffraction intensity modulations of the Bi$_2$Te$_3$ A$_{1g}^{(1)}$ mode are consistently about three times larger than in Bi$_2$Se$_3$.
Meanwhile, the (445) peak sensitivity to A$_{1g}^{(1)}$ distortion is about three times larger in Bi$_2$Te$_3$ than in Bi$_2$Se$_3$, due to their different eigenvectors. Therefore, the computed displacements $u_{\mathrm{XRD}}$ of A$_{1g}^{(1)}$ in the two materials are concluded to be similarly driven, see Fig.~\ref{fig:SACLA_amplitude} (b), where the slope of linear fit (broken lines) to Bi$_2$Te$_3$ is a factor of 0.88 lower than Bi$_2$Se$_3$.

The SS energy shifts in Bi$_2$Se$_3$ were measured with time-resolved ARPES under the same conditions as Bi$_2$Te$_3$, but with a slightly worse time resolution of 85 fs.
In Fig.~\ref{fig:SACLA_amplitude} (c) we show the SS energy shifts for both materials at similar absorbed fluences. As seen from the Fourier transforms in Fig.~\ref{fig:SACLA_amplitude} (d), Bi$_2$Se$_3$ has both A$_{1g}^{(1)}$ ($2.2$~THz) and A$_{1g}^{(2)}$ ($5.2$~THz)  phonons, similar to Bi$_2$Te$_3$. 
The frequencies of Bi$_2$Se$_3$ A$_{1g}$ phonons are higher than those in Bi$_2$Te$_3$ largely due to the lower average atomic mass. 
Analysis of the ARPES data shows that while the overall band shift is substantially larger in  Bi$_2$Se$_3$, the oscillatory component corresponding to the A$_{1g}^{(1)}$ mode in Bi$_2$Te$_3$ is a factor of 2.05 larger than in Bi$_2$Se$_3$ for the same fluence.

\section{Calculation of deformation potentials}

The section first discusses the calculations of Bi$_2$Te$_3$ DPs, which use numerical simulation to correct for intrinsic and extrinsic effects. 
We separately consider each effect below to provide the reader with intuition about the various ingredients and how to estimate their relative contributions.
We then discuss the calculations of Bi$_2$Se$_3$ DP, which is based on properly scaling the results of Bi$_2$Te$_3$ we established previously. 
    
\subsection{Calculations for Bi$_2$Te$_3$: Intrinsic effects}

     To compute the DP from experimental quantities, ideally, the measurements would be performed under identical pump conditions. For a given sample and pump wavelength, it means measurements would be performed at the same  fluence $F$. Experimentally, it is convenient to measure the linear $F$-dependence of the displacements and energy shifts, and use the fitted slopes to determine the DP:

    \begin{equation}\label{eq:definition of DP with F}
        \text{DP}= \frac{(d\Delta E_{\rm SS}/dF)}{(du_{\rm SS}/dF)}
    \end{equation}
    
    ARPES measures directly the SS energy shift ($\Delta E_{\rm SS}$), while XRD measures a weighted average of the atomic displacements over the probed region of the sample ($u_{\textrm{XRD}} \neq u_{\rm SS}$). Above we defined $u_{\rm SS}$ in Eq.~\ref{eq:definition of DP} as a weighted average over the SS wave function extension. 
    This brings uncertainties intrinsically related to the nature of the measurements which need to be addressed before Eq.~\ref{eq:definition of DP with F} can be applied to compute the DP. 
    To address this, we rescale the XRD measurements to give the atomic displacements in the vicinity of the SS:

        \begin{equation} \label{eq:rescale}
            u_{\rm SS} = \left(\frac{\tilde{u}_{\textrm{SS}}}{\tilde{u}_{\textrm{XRD}}}\right) u_{\textrm{XRD}}  
        \end{equation}
       \noindent 
       Quantities with tildes are obtained from simulations.
       The factor $\left(\frac{\tilde{u}_{\textrm{SS}}}{\tilde{u}_{\textrm{XRD}}}\right)$ is the simulated ratio between the average displacement $ \tilde{u}_{\textrm{SS}}$ over the spatial extent of the SS and the average displacement $\tilde{u}_{\textrm{XRD}}$ measured by XRD. 
       As we shall show, this ratio can be calculated by modeling the generation process of the depth-dependent coherent phonon field $u(z,t)$. For conceptual clarity, we begin by neglecting diffusion and recombination, and return to examine them later in this section.
 
   The depth-dependence of XRD is given by the x-ray penetration depth, which, for Bi$_2$Te$_3$ is 55~nm for our experimental condition (X-ray grazing incident at 0.5$^{\circ}$. In contrast, the SS wavefunction extends only $\sim$1~nm~ from the surface~\citep{pertsova2014probing}. 
   These lengthscales are denoted by $1/\alpha_i$, where ($i$=XRD) corresponds to the X-ray penetration depth at the incident angle in an XRD measurement, and ($i$=SS) corresponds to the spatial extent of the SS wave function. The change in scattering intensity in XRD is taken to be a weighted average of the phonon amplitude over the probed region. Similarly, the energy oscillation of the SS is given by displacements averaged over the extent of its wave function. These weighted averages can be written:
    
    \begin{equation}\label{eq:signal level}
\tilde{u}_{i}(t)=\frac{\int_{0}^{L_i} dz\exp{(-\alpha_i z)}u(z,t)}{\int_0^{L_i} dz\exp{(-\alpha_i z)}}
\end{equation}

\noindent where ${L_i}$ represents the thickness of the sample. To compute the displacements $u(z,t)$, we model $u(z,t)$ as a harmonic oscillator displaced to a new quasi-equilibrium coordinate $u'(z,t)$ \citep{zeiger1992}:

\begin{equation}\label{eq: phonon field}
    \frac{d^2 u}{dt^2}=-(2\pi f )^2 (u(z,t)- u'(z,t))-\gamma\dot{u}(z,t)
\end{equation}

\noindent where $f$ is the frequency and $\gamma$ is the damping rate. The displaced coordinate $u'(z,t)$ is proportional to the photoinduced carrier density $n(z,t)$, which has the initial condition:

\begin{equation}\label{init_excitation_carrier}
n(z,t=0) \propto \exp(-\alpha_0 z)
\end{equation} 

\noindent where $\alpha_0$ ($1/\alpha_0$=15~nm) is the absorption coefficient of the pump. Neglecting diffusion and recombination, the carrier density is static, and Eq.~\ref{init_excitation_carrier} is applicable for all times $t>0$, that is $n(z,t>0)=n(z,t=0)$. 
From Eq.~\ref{eq:signal level} and \ref{init_excitation_carrier} we obtain:
    \begin{equation}\label{eq:average carrier}
        \tilde{u}_{i}=u_0\frac{\alpha_i}{\alpha_i+\alpha_0}\frac{1-e^{-(\alpha_0+\alpha_i){L_i}}}{1-e^{-\alpha_i {L_i}}}
    \end{equation}
    where $u_0$ represents the displacement at $z=0$. We find that 
    the rescaling factor is $\left(\frac{\tilde{u}_{\textrm{SS}}}{\tilde{u}_{\textrm{XRD}}}\right) = 2.6$,
    which shows that XRD probes a deeper region in which the weighted average displacements are substantially attenuated compared to the surface. 
        
We now examine the possible role of carrier diffusion and recombination, which redistribute carriers following photoexcitation and can therefore introduce a time- and depth-dependence to the phonon driving forces. To motivate this discussion, we compute a back-of-the-envelope estimate of the relevance of diffusion: Using the ambipolar diffusivity of Bi$_2$Te$_3$ of $D_e =$ 520~nm$^2$/ps~\citep{jeon1991electrical}, we estimate a characteristic diffusion timescale of $1/(D_e \alpha_0^2) \approx 0.4$~ps. This is comparable with the phonon period for A$_{1g}^{(2)}$ of 0.25~ps. We conclude that diffusion redistributes carriers on a timescale comparable to that of the coherent phonon generation process, and its effect on the driving forces should be investigated.

To quantitatively model the effect, we retain Eq.~\ref{init_excitation_carrier} as an initial condition but allow the distribution to evolve via ambipolar diffusion along the sample normal direction $\hat{z}$:
\begin{equation}\label{diffusion}
\frac{dn}{dt}=D_e \frac{d^2n}{dz^2}-\frac{1}{\tau_0} n
\end{equation}
where $\tau_0$ is the carrier lifetime, which is estimated to be 20~ps~\citep{sheu2013free}. We apply a boundary condition
\begin{equation}\label{boundary condition}
\frac{dn}{dz}(z=0)=0, \frac{dn}{dz}(z=L_i)=0
\end{equation} 
to enforce no diffusion across the sample boundaries at 0 and $L_i$.

 We obtain the carrier density profile $n(z,t)$, by solving Eq.~\ref{diffusion}. To connect this to our experiment, we can then compute the coherent phonon amplitude  $u(z,t)$ from Eq.~\ref{eq: phonon field} by taking $u'(z,t) \propto n(z,t)$ as before, but now with the time-dependence included. However, for the most faithful modeling of the experiment, we must account for the fact that the phonon frequency is itself a function of $n(z,t)$, as demonstrated experimentally by fluence-dependent mode softening in Fig.~\ref{fig:diffusion3}(a).  We model this as a linear dependence of the mode frequency $f(z,t)$ on fluence: $f(z,t)=f_0 (1-\beta n(z,t))$. Here  $f_0$ is the equilibrium frequency and 
$\beta > 0 $ is the proportionality constant which determines the magnitude of the softening. With this assumption, we can numerically solve for $u(z,t)$, and finally calculate the average displacements $\tilde{u}_i(t)$ using Eq.~\ref{eq:signal level}.

Here we take the A$_{1g}^{(2)}$ mode as an example, and describe the simulation of Bi$_2$Te$_3$, assuming $D_e$=520~nm$^2$/ps \citep{jeon1991electrical}.
We choose $\beta$ to reproduce the observed linear softening, as shown in Appendix C.
The damping constant $\gamma$ is taken to be 0.25~ps$^{-1}$, to be consistent with the experimental data. We find that $\left(\frac{\tilde{u}_{\textrm{SS}}}{\tilde{u}_{\textrm{XRD}}}\right) = 2.2$ for A$_{1g}^{(2)}$. The corresponding factor for the A$_{1g}^{(1)}$ mode is 2.0.
These can be compared with the factor of 2.6 we computed earlier from Eq.~(\ref{eq:average carrier}), which neglected diffusion and recombination. This indicates that the carrier dynamics introduce a $20-30$\% correction on top of the penetration depth considerations. Therefore, for quantitative purposes it is necessary to account for diffusion and recombination; however, the dominant correction arises from the large penetration depth of the XRD measurement compared to the extent of the SS wave function.

\subsection{Calculations for Bi$_2$Te$_3$: Extrinsic effects}

We now return to the issue of extrinsic effects. The first is the experimental uncertainty of fluence measurements in different experimental facilities. The second is the correction required by different time resolutions $\tau$ in ARPES and XRD measurements. 

The precision and accuracy of fluence measurements can vary substantially between different labs: in published Bi measurements, the reported fluences for the same amount of phonon softening differ by up to a factor of 5 in independent measurements\citep{boschetto2008small,wu2007coupling,misochko2004observation,decamp2001dynamics}. These discrepancies may be attributed to miscalibration in the laser power measurement or errors in the measurement of the incident area on the sample. Therefore, for quantitative calculation of the DP, we investigate fluence uncertainties between XRD and ARPES measurements.

Fluence uncertainties can be calibrated by using the A$_{1g}^{(2)}$ frequency as a proxy due to its strong dependence on fluence, as shown in Appendix C. 
Explicitly, we need to define
\begin{equation}\label{eq:EXP definition of uSS 2}
u_{\rm SS} = \frac{(d\tilde{u}_{\textrm{SS}}/d\tilde{f}_{\textrm{SS}})}{(d\tilde{u}_{\textrm{XRD}}/d\tilde{f}_{\textrm{XRD}})} u_{\textrm{XRD}}
\end{equation}
in place of Equation ~\ref{eq:rescale}, and accordingly 
\begin{equation}\label{eq:EXP definition of DP fixed freq 2}
        \text{DP}= \frac{(d\Delta E_{\textrm{SS}}/df_{\textrm{SS}})}{(du_{\rm SS}/df_{\textrm{XRD}})}
\end{equation}
\noindent where $f_i$ is the A$_{1g}^{(2)}$ frequency measured by the respective techniques. 
We note that $f_{\textrm{SS}} \neq f_{\textrm{XRD}}$ even for the same incident fluence $F$. 
This is again due to the depth-dependence of phonon fields, and the different sensitivities of ARPES and XRD as described by Eq.~\ref{eq:signal level}.
Eq.~\ref{eq:EXP definition of DP fixed freq 2} is mathematically equivalent to the original definition (Eq.~\ref{eq:definition of DP with F}). By doing so, uncertainties in fluence calibration are effectively circumvented by using the softened frequency as a proxy. 
The calculation of $(d\tilde{u}_{\textrm{SS}}/d\tilde{f}_{\textrm{SS}})/(d\tilde{u}_{\textrm{XRD}}/d\tilde{f}_{\textrm{XRD}})$ is detailed in Appendix C.

\begin{table}
		\begin{ruledtabular}
			\begin{tabular}{c|c|c}
				   & experiment & theory (LDA)\\
				\hline
				$\rm{Bi_2Te_3}$, A$_{1g}^{(1)}$   & 2.4 (0.7) & 5.8\\
				$\rm{Bi_2Te_3}$,A$_{1g}^{(2)}$  & 2.7 (0.8) & 6.8\\
				$\rm{Bi_2Se_3}$, A$_{1g}^{(1)}$  & 1.8 (0.7) & 1.2\\
			\end{tabular}
		\end{ruledtabular}
		\caption{\label{tab:Deformation potential} Deformation potential [meV/pm]. The experimental values are averaged over the region of interest across $\textbf{k}_{\vert\vert}$ space as shown in Fig.~\ref{fig:XRD&ARPES} (d). 
		The experiment values account for intrinsic and extrinsic corrections. 
        Error bars include both systematic and statistical error.
        Theory values are taken at the $\Gamma$-point.  
		 }
\end{table}

	\begin{figure} 
		\includegraphics{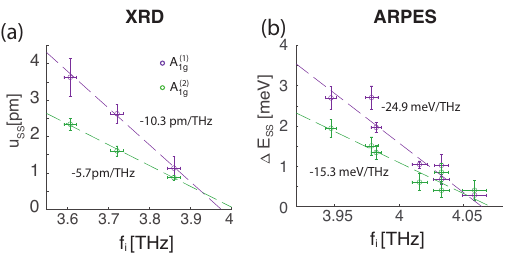}
		\caption{ 
        A$_{1g}^{(2)}$ frequency dependence in Bi$_2$Te$_3$. 
		(a) Displacement $u_{\textrm{SS}}$ and (b) SS band shift $\Delta E_{\textrm{SS}}$ as a function of the measured A$_{1g}^{(2)}$ frequency.}
		\label{fig:diffusemain}
	\end{figure}
 
 Both measured quantities $u_{\rm XRD}$ and $\Delta E_{\textrm{SS}}$ need to be multiplied by an additional correction factor due to limited time resolution $\tau$ (60~fs in ARPES  and 50~fs in XRD measurements)~\cite{merlin1997generating}:
 \begin{equation}\label{eq:damping factor}
    \exp{(\frac{\pi^2 f_0^2 \tau^2}{4 \text{ln}2})}
\end{equation}
where $f_0$ represents A$_{1g}$ mode frequency $f_0$.

Eq.~\ref{eq:EXP definition of DP fixed freq 2} is then used to calculate DPs of A$_{1g}$ modes in Bi$_2$Te$_3$. 
In Fig.~\ref{fig:diffusemain} (a), we plot the displacement $u_{\rm SS}$ and the electronic band shift $\Delta E_{\textrm{SS}}$ as a function of A$_{1g}^{(2)}$ frequency. 
The slopes of these linear fits are used to calculate the DPs in Eq.~\ref{eq:EXP definition of DP fixed freq 2}.
The resulting DPs are listed in Table I, which reflects our best estimate considering all intrinsic and extrinsic factors. 
We note that these values differ by 20\% compared with the DP calculation results without using frequency as a proxy for fluence, suggesting a fluence mismatch of 20\% between the two measurement facilities.

The error bars in the table are estimated as follows.
Statistical errors are obtained from the fitting error bars of Fig.~\ref{fig:diffusemain} (a,b) which weighs in both the horizontal error bars and vertical error bars. 
Systematic errors are estimated based on the uncertainty in two important parameters of our calculation: the x-ray incidence angle (and hence the penetration depth), and the magnitude of the diffusivity. We estimate a $\pm$0.1$^{\circ}$ uncertainty with respect to the nominal incidence angle 0.5$^{\circ}$, which results in penetration depths ranging from 35~nm to 70~nm. This introduces an error of 8\% into DP calculation. 
To estimate the systematic error associated with diffusivity, we varied the value from 200nm$^2$/ps to 1000nm$^2$/ps, which differ from the estimated value of $D_e$=520nm$^2$/ps by roughly a factor of two. This introduces a corresponding  15\% uncertainty  in the DP.  

\subsection{Calculation for Bi$_2$Se$_3$}

To compute the DP of Bi$_2$Se$_3$, we cannot directly employ the strategies from the previous section II.B  that minimize the systematic errors due to fluence calibration. 
The A$_{1g}^{(2)}$ mode of Bi$_2$Se$_3$ was not observed by XRD, and is therefore not available as a fluence proxy.
Instead, we exploit that Bi$_2$Se$_3$ and Bi$_2$Te$_3$ were measured side-by-side by XRD and thus without systematic fluence uncertainty. This allows us to compute the ratio of their DPs as defined in Eq.~\ref{eq:rescale} and Eq.~\ref{eq:definition of DP with F}:
\begin{equation}\label{eq:redefinition of DP with F}
        \text{DP}= \left(\frac{\tilde{u}_{\textrm{SS}}}{\tilde{u}_{\textrm{XRD}}}\right)^{-1}\frac{(d\Delta E_{\rm SS}/dF)}{(du_{\rm XRD}/dF)}
    \end{equation}

From the XRD measurements at SACLA and the ARPES measurements as shown in Fig.~\ref{fig:SACLA_amplitude}, we obtain the ratio 2.33 of the quantity $\frac{(d\Delta E_{\rm SS}/dF)}{(du_{\rm XRD}/dF)}$ between Bi$_2$Te$_3$ and Bi$_2$Se$_3$. 

We then need to take into account that the different ambipolar diffusivity in Bi$_2$Se$_3$ is much higher $D_e=$1400nm$^2$/ps~\citep{Thermoelectric2009}. Diffusion, therefore, leads to a lower carrier density in the vicinity of the SS in Bi$_2$Se$_3$ as compared to Bi$_2$Te$_3$, leading to a lower $\tilde{u}_{\textrm{SS}}$, and thus an additional correction factor of 0.63 to the ratio of DPs between Bi$_2$Te$_3$ and Bi$_2$Se$_3$, due to the quantity $\left(\frac{\tilde{u}_{\textrm{SS}}}{\tilde{u}_{\textrm{XRD}}}\right) $. See Appendix C for more details. Lastly, we account for the correction due to time resolution differences, which introduces a factor of 0.93. We arrive at the final DP ratio of Bi$_2$Te$_3$ over Bi$_2$Se$_3$ of 1.37. 

Finally, knowing the A$_{1g}^{(1)}$ DP ratio we can calculate the absolute value Bi$_2$Se$_3$ DP based on our knowledge of the Bi$_2$Te$_3$ DP value, see table~\ref{tab:Deformation potential}.
The uncertainties include statistical error (32\%), as well as systematic error (16\%) from uncertainties in X-ray incidence angle calibration and ambipolar diffusivities. 

We also estimate a lower bound of the Bi$_2$Se$_3$ A$_{1g}^{(2)}$ DP of 0.2~meV/pm. This bound is based on the non-observation of this mode in the SACLA XRD measurement, where its amplitude was below the noise level. 

\begin{figure}
		\includegraphics{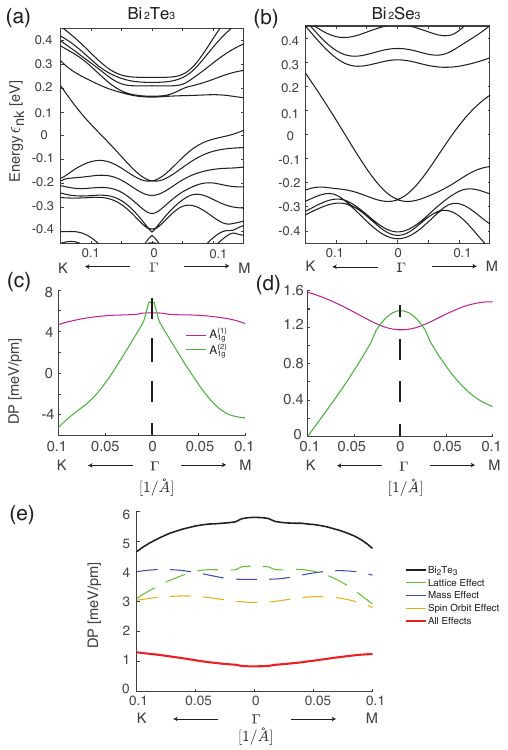}
		\caption{DFT calculation of electronic band structures and deformation potentials. (a) Band structure for a 6-quintuple layer (QL) slab of Bi$_2$Te$_3$, shown for part of the reciprocal space.  
		(b) Band structure for a 5-QL slab of Bi$_2$Se$_3$.  
		(c) Deformation potential (DP) of the two A$_{1g}$ modes in Bi$_2$Te$_3$ derived from  frozen phonon calculations. 
	    (d) Deformation potentials of the A$_{1g}$ modes in Bi$_2$Se$_3$.  
	    (e) Calculated A$_{1g}$ mode DP by swapping the lattice constants value, masses of group-VI atoms, and spin-orbit coupling strength in Bi$_2$Te$_3$ with that of Bi$_2$Se$_3$. Combining all three effects, the prediction of factor 4.8 larger A$_{1g}$ DP in Bi$_2$Te$_3$ can be reasonably explained.}
		\label{fig:theory}
\end{figure}

\section{DFT calculations}
	
We performed DFT calculations to compare with the experimentally measured A$_{1g}$-SS DPs of Bi$_2$Te$_3$ and Bi$_2$Se$_3$. The details of our calculations are given in Appendix E. The computed electronic band structures of Bi$_2$Te$_3$ and Bi$_2$Se$_3$ slabs are shown in Fig.~\ref{fig:theory} (a) and (b).
DPs are obtained using the frozen phonon method, where the atoms are displaced along the A$_{1g}$ mode coordinates, and the SS energy shifts are calculated based on the resulting shifts in the electronic band structure of a slab. 
In our DP calculations, we use the vacuum level of the Hartree potential to align the energies of the electronic states. We assume that bulk phonon modes are homogeneously distributed along the slab despite the surface's presence. 
The computed DPs along K-$\Gamma$-M are shown in Fig.~\ref{fig:theory} (c) and (d).

We select the computed DP values at the $\Gamma$-point to compare with the experimental values, and include them in Table \ref{tab:Deformation potential}. 
The calculated A$_{1g}^{(1)}$ DP in Bi$_2$Se$_3$ is 1.5 times smaller than in the experiment and agrees within the experimental uncertainties.
In Bi$_2$Te$_3$, the A$_{1g}$ DPs are about 2.5 times larger than the experimental values and outside of the estimated confidence interval of measurements.
We emphasize that this comparison does not reveal an order of magnitude difference as found in the strongly correlated superconductor FeSe~\cite{gerber2017femtosecond} and that our values are roughly one order of magnitude smaller than those
for other non-topological semiconductors such as Si and Ge ~\cite{Fahy2008}. 
Considering the theoretical and experimental difficulties in obtaining precise and accurate values on a scale of a few meV per unit cell, we regard a difference of factor 2 as general agreement.

From the perspective of theory, it is interesting that calculations for the A$_{1g}^{(1)}$ mode predict a factor 4.8 larger DP in Bi$_2$Te$_3$ than in Bi$_2$Se$_3$. This only agrees qualitatively with the experiment which observed a factor 1.4 larger DP in Bi$_2$Te$_3$.
This quantitative discrepancy might be attributed to the limitations of DFT calculations in capturing the precise band structure.
We note that in the experiment, the A$_{1g}^{(1)}$ DPs are weakly momentum-dependent near the zone center (see Appendix B), whereas theory predicts strong momentum-dependence of the A$_{1g}^{(2)}$ DPs.
Calculations that account for electron correlations may result in a better agreement with the experimentally measured band structure, as suggested by ref~\citep{forster2016}, as well as our DP measurements.


To understand the difference in the DP values of Bi$_2$Te$_3$ and Bi$_2$Se$_3$ computed with DFT,
we conducted a numerical experiment in which we modified individual parameters that are different in the two systems. 
These are: lattice constants, Se and Te atomic masses, and the spin-orbit coupling strength. 
This thought experiment sheds light on the individual contributions of these parameters to the DPs in both materials.

We start with Bi$_2$Te$_3$ and evaluate the contribution of each individual parameter of Bi$_2$Se$_3$ to the DPs of Bi$_2$Te$_3$. 
Note that we only varied one parameter at a time, while holding the rest at their nominal values for Bi$_2$Te$_3$. The resulting A$_{1g}^{(1)}$ deformation potentials are plotted in Fig. \ref{fig:theory} (e). 
If the Se atomic mass replaces the Te atomic mass, the DP is reduced by $\sim 35\%$ due to the change in eigenvector (see Table ~\ref{tab:Eigenvector}). 
By modifying the spin-orbit interaction strength of Te to the value in Se, the DP is reduced by $\sim 50\%$.
When the Bi$_2$Se$_3$ lattice parameters are used for the Bi$_2$Te$_3$ system, the DP at $\Gamma$ is reduced by $\sim 27\%$, which is mostly related to the lattice expansion affecting the crystal field splitting. 
Finally, when all these factors are simultaneously incorporated, Bi$_2$Te$_3$ is effectively transformed into Bi$_2$Se$_3$, reproducing the factor $\sim 5$ difference of native LDA. Our numerical experiment demonstrates that no single factor is sufficient to explain the difference in the calculated DPs of Bi$_2$Te$_3$ and Bi$_2$Se$_3$, and that lattice constants, atomic masses, and spin-orbit coupling strength are all key ingredients.

Additionally, we performed a constrained DFT calculation (see Appendix E) of the vibrational modes for the photoexcited bulk Bi$_2$Te$_3$ and Bi$_2$Se$_3$. We found that for Bi$_2$Te$_3$, the mode amplitude dependence on A$_{1g}^{(1)}$ frequency reproduces the measurement well, and that the ratio between Bi$_2$Te$_3$ and Bi$_2$Se$_3$ phonon amplitudes given the same excitation density is consistent with the experiment.  
  
\section{Conclusions}
We combined time-resolved XRD and ARPES measurements on the topological insulators Bi$_2$Te$_3$ and Bi$_2$Se$_3$ and quantify the coupling between topological surface states and A$_{1g}$ phonon modes by extracting mode- and band-resolved deformation potentials.
We compare experimental results to DFT and find reasonable agreement.
Our results directly impact the microscopic understanding of topological SS transport and topological phase transitions\citep{aryal2021topological}. 
Our methodology lays the systematic groundwork required for future work studying electron-phonon coupling in more complex quantum materials.

For the dominant A$_{1g}^{(1)}$ phonon mode, we find a DP on the order of $2$~meV/pm in both materials. 
The coupling is 1.4 times larger in Bi$_2$Te$_3$ than in Bi$_2$Se$_3$, yet this difference is insignificant within experimental uncertainties. 
Our DFT calculations agree with the experimental findings within a factor of 2.5, which we regard as agreement considering the small energy and length scales involved.
While we can not exclude that correlation effects in the calculations might improve agreement with the experiment, we did not observe an order of magnitude discrepancy as was revealed for the strongly correlated superconductor FeSe~\cite{gerber2017femtosecond}. 
In addition, excited state DFT correctly predicts the experimentally observed mode softening. 
We also analyzed the individual contributions of lattice constants, atomic masses, and spin-orbit coupling strength to the difference between the DFT values of DPs for Bi$_2$Te$_3$ and Bi$_2$Se$_3$.
All parameters contribute considerably, with spin-orbit coupling providing the largest contribution.
These favorable results underline that DFT in this material class is well capable of predicting how photoexcited electrons couple to phonon modes and of capturing excited potential energy surfaces.

We performed a comprehensive analysis of correction factors due to intrinsic and extrinsic experimental effects. 
The most substantial intrinsic effect is the different depth-sensitivity of the two probes, which leads to a different mismatch of pumped and probed sample volumes. 
Only smaller corrections are necessary to account for photoexcited carrier diffusion and recombination.  
Extrinsic effects are dominated by uncertainties of the excitation density. 
We were able to circumvent this extrinsic issue by using a strongly fluence dependent phonon mode frequency as inherent calibration.

Our insights hold important positive lessons for future work. 
The strongest intrinsic effect of different probing depths can be greatly reduced by experiments on films thinner than the pump's optical penetration depth. 
This ensures a homogeneous distribution of excited carriers regardless of the probe penetration depth and also mitigates diffusion effects. 
This approach taken in Ref.~\cite{gerber2017femtosecond} results in only a modest factor of 1.03 to account for the different penetration depths of ARPES and XRD.

Looking further ahead, we suggest using higher and tunable photon energies for time-resolved ARPES. 
This will allow accessing the dynamics of bulk states in well-defined conditions, which this work was not able to accomplish.
We expect that future multi-modal experiments will combine time-resolved ARPES and diffraction methods in a single instrument. 
Most if not all extrinsic effects will be eliminated in such simultaneous experiments. 
Free electron laser facilities will provide this platform to open a new experimental paradigm for studying electron-phonon coupling with high accuracy and precision.

\begin{acknowledgments}
Y.H., J.A.S., S.T., M.T., P.S.K., T.H., M.J., Z-X.S and D.A.R.  were supported by the U.S. Department of Energy, Office of Science, Office of Basic Energy Sciences through the Division of Materials Sciences and Engineering under Contract No. DE-AC02-76SF00515. 
Use of the LCLS and SSRL is supported by the US Department of Energy, Office of Science, Office of Basic Energy Sciences under Contract No. DE-AC02-76SF00515. The SACLA experiment was performed with the approval of the Japan Synchrotron Radiation Research Institute (JASRI; Proposal No. 2018B8079). T. K. acknowledges JSPS KAKENHI (Grant Numbers JP19H05782, JP21H04974,and JP21K18944).
The DFT work was supported by Science Foundation Ireland and the Department for the Economy Northern Ireland Investigators Program under Grant Nos. 15/IA/3160 and 12/IA/1601.
Preliminary x-ray characterization was performed at beamline 7-2 at the Stanford Synchrotron Radiation Lightsource (SSRL).
The ellipsometry measurement and thin film reflection measurement were performed at the Stanford Nano Shared Facilities (SNSF), supported by the National Science Foundation under award ECCS-2026822. 
Y.H. acknowledges help with SACLA beamtime from Christopher P. Weber, Matthew J. Kim, and Ka Lun Michael Man. Y.H. and D.A.R. also thank Christopher P. Weber for suggestions on the manuscript.
\end{acknowledgments}

\appendix
\section{Crystal structures}

\begin{figure} 
		\includegraphics{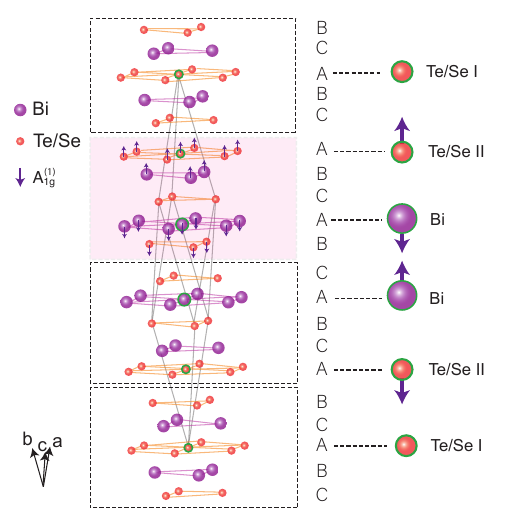}
		\caption{Crystal structure of Bi$_2$X$_3$ (X = Se,Te) and schematic for A$_{1g}^{(1)}$ phonon mode.
		\label{fig:crystal}}
\end{figure}

Fig.~\ref{fig:crystal} shows the crystal structure of space group R$\bar{3}$m, and provides a 1D-atom-chain view of A$_{1g}^{(1)}$ mode on the right side. The quintuple-layer view of A$_{1g}^{(1)}$ mode as Fig.~\ref{fig:XRD&ARPES} (a) is provided in the pink box. 
The layer pattern is represented with A, B and C. Atoms forming the 1D-atom-chain are all taken from the A layer. 
The primitive unit cell is outlined with gray solid lines. 
Among the three types of atoms participating in A$_{1g}$ modes, Bi, X(I), and X(II), the X(I) atoms are not allowed to displace in a A$_{1g}$ mode as these atoms are the centers of inversion symmetry. 
Thus, an A$_{1g}$ phonon has only two degrees of freedom determined by Bi and X(II) atomic motion along [111]. Since there are two free parameters, there are correspondingly two A$_{1g}$ modes. 

First, we refine the internal atomic coordinates using the static diffraction intensity on different peaks. Second, to determine the eigenvector, we use the diffraction intensity modulation induced by the same phonon mode measured across multiple Bragg peaks. Finally, based on the measured eigenvectors, $u^{\mathrm{Bi}}_{\lambda}$ can be quantified from diffraction. We detail the second step below.

We parameterize the X(II) atom coordinates in the primitive unit cell as $(\delta_1,\delta_1, \delta_1)$ and the Bi atom coordinates as $(\delta_2,\delta_2, \delta_2)$. Then the scattering structure factor is given by
\begin{equation} 
\begin{split}
    F =f_\text{X}+2e^{i\pi (h+k+l)}[f_\text{X}\cos[\pi(1-2\delta_1)(h+k+l)]\\
    +f_\text{Bi}\cos[\pi(1-2\delta_2)(h+k+l)]]
\end{split}
\end{equation}
with $\delta_1 \approx \frac{1}{5}$ and $\delta_2 \approx \frac{2}{5}$. For Bragg peaks, $h,k,l$ are integers, and we define $h+k+l=N$.
For $N=5n$, the Bragg peaks are not sensitive to either of the A$_{1g}$ modes. 
For $N=5n\pm1$, peaks have the same sensitivity to a given A$_{1g}$ mode for a given $n$, and are sensitive to both A$_{1g}^{(1)}$ and A$_{1g}^{(2)}$. 
For $N=5n\pm2$, peaks have the same sensitivity to a given A$_{1g}$ mode for a given $n$, and are sensitive mainly to the A$_{1g}^{(1)}$ mode.

During the XRD experiment at LCLS we oriented the sample by rotating it around the azimuthal axis, which was aligned to the sample surface normal to within 0.1$^{\circ}$. We so obtained the static X-ray scattering intensity on a total of seven peaks. This allows us to refine the lattice constants and the internal coordinates ($\delta_1, \delta_2$), which are in good agreement with DFT calculations.

	\begin{figure} 
		\includegraphics{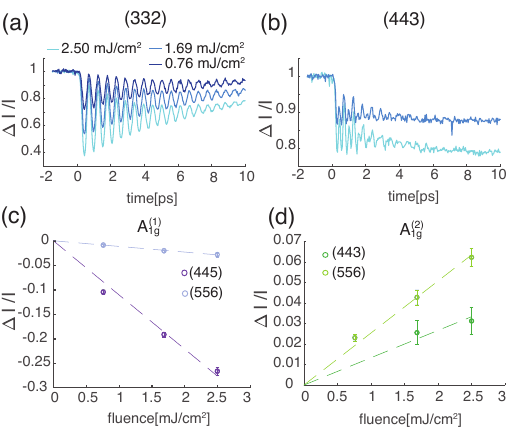}
		\caption{Bi$_2$Se$_3$ XRD data from LCLS for the eigenvector reconstruction. (a) (b) Time-resolved diffraction on peaks (332) and (443).  Peak intensity modulation amplitude induced by a given A$_{1g}$ mode is different on peaks (332) and (443), which contains information for calculating the A$_{1g}$ eigenvector. (c) A$_{1g}^{(1)}$ induced normalized intensity modulation amplitude $\Delta I/I$ on peaks (445) and (556). (d) A$_{1g}^{(2)}$ induced intensity modulation amplitude on peaks (443) and (556).
		}
		\label{fig:LCLS}
	\end{figure}

\begin{table}
		\begin{ruledtabular}
			\begin{tabular}{c|c|c}
			\centering
				mode & Bi & Te/Se II \\
				\hline
				$\rm{Bi_2Te_3}$, A$_{1g}^{(1)}$  & 0.44 & 0.56\\
				$\rm{Bi_2Te_3}$,A$_{1g}^{(2)}$  & 0.56 & 0.44\\
				$\rm{Bi_2Se_3}$, A$_{1g}^{(1)}$  & 0.58  & 0.40\\
				$\rm{Bi_2Se_3}$, A$_{1g}^{(2)}$  & 0.40  & 0.58\\
			\end{tabular}
		\end{ruledtabular}
		\caption{\label{tab:Eigenvector} The normalized eigenvectors of $\rm{Bi_2Te_3}$ and $\rm{Bi_2Se_3}$ A$_{1g}$ modes. Bi$_2$Te$_3$ eigenvectors are obtained using XRD, while those of Bi$_2$Se$_3$ are calculated using DFT. }
	\end{table}
 
The eigenvector of Bi$_2$Te$_3$ A$_{1g}$ phonon mode can be obtained from the intensity modulation amplitude of a set of peaks sensitive to the mode. 
We need only one free parameter to describe two A$_{1g}$ mode eigenvectors simultaneously, and we pick this free parameter to be the Bi/Te ratio in the A$_{1g}^{(1)}$ mode.
The experimental observables are the Bragg peak intensity modulations induced by an A$_{1g}$ mode (Fig.~\ref{fig:LCLS} (a) and (b)). As shown in Fig.~\ref{fig:LCLS} (c) and (d), amplitudes of the normalized Bragg peak intensity modulation $\Delta I/I$ linearly increase with fluence. The increase in $\Delta I/I$ per unit fluence (1~mJ/cm$^2$) is the slope of the linear fit constrained through the origin in Fig.~\ref{fig:LCLS} (c) and (d).

We aim to solve for three unknown variables: (1) the Bi/Te ratio in the A$_{1g}^{(1)}$ mode, (2) the A$_{1g}^{(1)}$ mode amplitude (alternatively one can also choose to use $u^{\mathrm{Bi}}$) increase per unit fluence, and (3) the A$_{1g}^{(2)}$ mode amplitude increase per unit fluence. 
We therefore define a set of three constraining equations: the linear slopes of (1) the A$_{1g}^{(1)}$ modulated diffraction intensity of the (556) peak versus fluence, (2) the A$_{1g}^{(1)}$ modulated diffraction intensity of the (445) peak versus fluence, and (3) the A$_{1g}^{(2)}$ modulated diffraction intensity of the (556) peak versus fluence.
To avoid near-degeneracy of the equation set, we pick two peaks that either satisfy $N=5n\pm1$ or $N=5n\pm2$ for the A$_{1g}^{(1)}$ mode. 
The solution of this equation set agrees with the eigenvectors calculated by DFT to within 15\%, when comparing the \emph{normalized} A$_{1g}^{(1)}$ eigenvector projected onto a Bi atom. 

Table ~\ref{tab:Eigenvector} shows the normalized eigenvectors of $\rm{Bi_2Te_3}$ calculated through XRD measurements, and of $\rm{Bi_2Se_3}$ obtained through DFT. For Bi$_2$Se$_3$ we did not resolve signals on a sufficient number of Bragg peaks to directly solve the eigenvectors. 

\section{$\textbf{k}$-dependence of SS energy shifts}
\begin{figure*} 
		\includegraphics{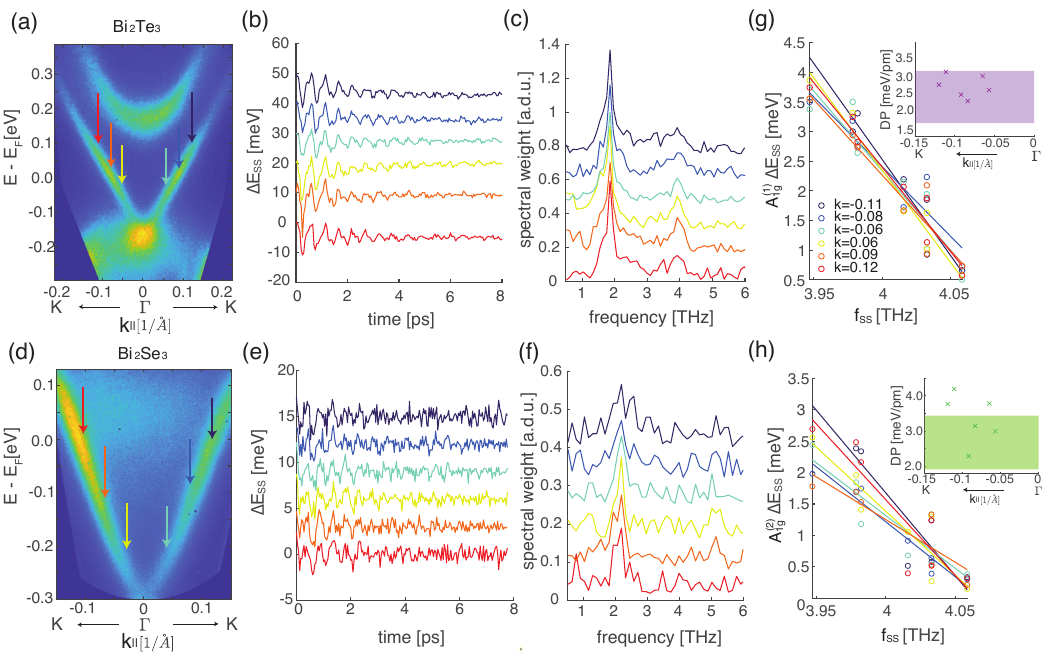}
		\caption{$\textbf{k}$-dependence of time-resolved ARPES measurements. 
		(a) The electron population of Bi$_2$Te$_3$ measured at 510~fs delay at 1.02~mJ/cm$^2$. The arrows show different $\textbf{k}_{\vert\vert}$ regions we sample for the SS. 
		(b) Modulation of Bi$_2$Te$_3$ SS at 1.02~mJ/cm$^2$, sampled at different $\textbf{k}_{\vert\vert}$ regions. 
		(c) The Fourier transform of (b).
		(d) The electron population of Bi$_2$Se$_3$ measured at 510~fs delay at 0.54~mJ/cm$^2$. The arrows show different $\textbf{k}_{\vert\vert}$ regions we sample for the SS. 
		(e) Modulation of Bi$_2$Se$_3$ SS at 0.54~mJ/cm$^2$, sampled at different $\textbf{k}_{\vert\vert}$ regions. 
		(f) The Fourier transform of (e).
		(g) Bi$_2$Te$_3$ A$_{1g}^{(1)}$ phonon induced energy modulation amplitude as a function of A$_{1g}^{(2)}$ frequency. The slopes of the linear fits across $\textbf{k}_{\vert\vert}$ points do not vary much, which justifies averaging over a large region of interest in $\textbf{k}_{\vert\vert}$ space to improve the signal-to-noise ratio. The inset contains part of Fig.~\ref{fig:theory} (c), with additional crosses representing the A$_{1g}^{(1)}$ DPs calculated for individual $\textbf{k}_{\vert\vert}$ points. (h) Bi$_2$Te$_3$ A$_{1g}^{(2)}$ phonon induced energy modulation amplitude as a function of A$_{1g}^{(2)}$ frequency. Crosses in the inset represent the A$_{1g}^{(2)}$ DPs calculated for individual $\textbf{k}_{\vert\vert}$ points.
		}
		\label{fig:ARPES_kdependence}
\end{figure*}

To enhance the signal-to-noise ratio of the surface electron band coherent oscillations, we average the change in band energy over $\pm$ 0.2 \AA~$^{-1}$ near $\Gamma$, which is based on the fact that the DPs and thus the electron band shifts are nearly constant across this area of reciprocal space. In Fig.~\ref{fig:ARPES_kdependence} we show the $\textbf{k}$-dependent SS response, the sampled $\textbf{k}_{\vert\vert}$ points are illustrated in Fig.~\ref{fig:ARPES_kdependence} (a) and (d). In Fig.~\ref{fig:ARPES_kdependence} (g) and (h), the energy shift versus A$_{1g}$ frequencies for different $\textbf{k}_{\vert\vert}$ points are plotted for A$_{1g}^{(1)}$ and A$_{1g}^{(2)}$ in Bi$_2$Te$_3$. 
The $\textbf{k}_{\vert\vert}$ dependence of A$_{1g}^{(1)}$-induced binding energy oscillations $\Delta E^{n\textbf{k}}$ is weak with a variation of 13\%.
The A$_{1g}^{(2)}$ has a lower signal-to-noise ratio, adversely impacting the amplitude extraction and leading to higher variations in DPs. 
Overall, averaging over a larger portion of $\textbf{k}$-space to improve the signal-to-noise ratio is justified. 

\section{Complete deformation potential calculation}

	\begin{figure} 
		\includegraphics{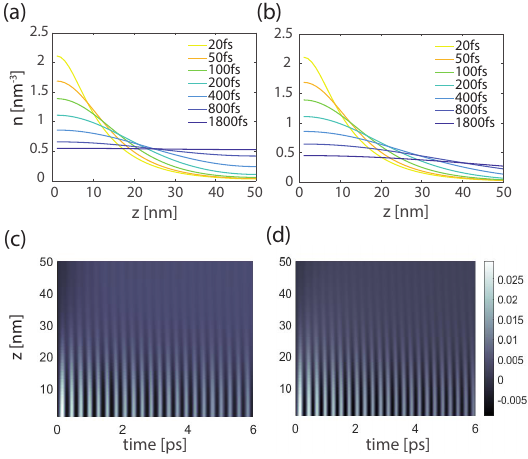}
		\caption{ Diffusivity simulation. 
		Nonequilibrium carrier density profile for different time delays after the laser pump of fluence 2mJ/cm$^2$, in a 50~nm thin film (a) and bulk (b). 
		Coherent phonon field induced by the population of nonequilibrium carriers, in a 50~nm thin film (c) which corresponds to (a), and bulk (d) which corresponds to (b). 
		}
		\label{fig:diffusion}
	\end{figure}
		
	\begin{figure} 
		\includegraphics{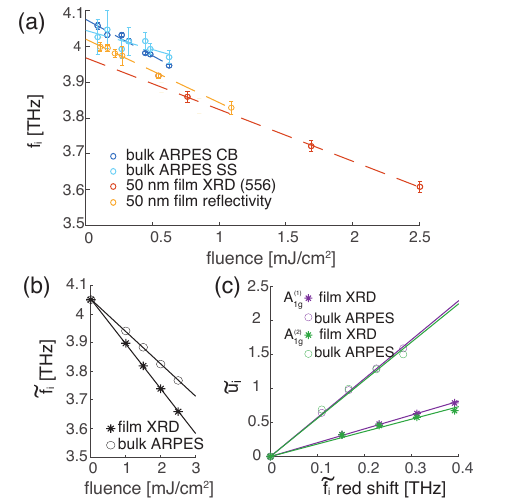}
		\caption{ (a) Linear fluence dependence of the Bi$_2$Te$_3$A$_{1g}^{(2)}$ mode frequency. We include data taken with XRD (50~nm film), ARPES (bulk) and pump-probe optical reflectivity (50~nm film), which span different ranges of absorbed fluences and thus different ranges of $f_i$.
		(b)  Simulated A$_{1g}^{(2)}$ mode frequency dependence on fluence. 	
		(c) Simulated A$_{1g}$ normalized signals of time-resolved ARPES of a bulk sample, and the normalized signals of time-resolved XRD for a 50~nm thin film sample used in the LCLS experiment as a function of A$_{1g}^{(2)}$ frequency red shift.}
		\label{fig:diffusion3}
	\end{figure}
	
As in the main text, we present the simulation details for DP calculations in  Bi$_2$Te$_3$ first, before we do so for Bi$_2$Se$_3$.  
We show the simulated carrier density $n(z,t)$ and the phonon field $u(z,t)$ in Fig.~\ref{fig:diffusion}, for both the thin film used in XRD and the bulk sample used in ARPES. 
This simulation is used to obtain the scaling factors in Eq.\ref{eq:rescale} and Eq.~\ref{eq:EXP definition of uSS 2} .

We show the measured fluence dependence of the mode frequencies $f_i$ in Fig.~\ref{fig:diffusion3} (a), for different samples and experimental conditions. The large change of the Bi$_2$Te$_3$ A$_{1g}^{(2)}$ frequency with increasing fluence shows that $f_i$ can be a good proxy for the absorbed pump energy density.
Thus, we can use Eq.~\ref{eq:EXP definition of DP fixed freq 2} to correct for the fluence uncertainty in different experiments. 

The DPs are proportional to $[\frac{(d\tilde{u}_{\textrm{SS}}/d\tilde{f}_{\textrm{SS}})}{(d\tilde{u}_{\textrm{XRD}}/d\tilde{f}_{\textrm{XRD}})}]^{-1}$, and hence 

\begin{equation}\label{eq:first two factors}
        [\frac{(d\tilde{u}_{\textrm{SS}}/d\tilde{f}_{\textrm{SS}})}{(d\tilde{u}_{\textrm{XRD}}/d\tilde{f}_{\textrm{XRD}})}]^{-1}=\left(\frac{\tilde{u}_{\textrm{SS}}}{\tilde{u}_{\textrm{XRD}}}\right)^{-1}\times \frac{(d\tilde{f}_{\textrm{SS}}/dF)}{(d\tilde{f}_{\textrm{XRD}}/dF)}
\end{equation}

The first factor in Eq.~\ref{eq:first two factors} right hand side is presented in the main text, assuming a fixed fluence. 
Simulated fluence ($F$) dependence of $\tilde{f}_i$ in XRD (using thin film) and ARPES (using bulk) is shown in Fig.~\ref{fig:diffusion3} (b). 
$\tilde{f}_i$ is obtained from $\tilde{u}_{i}(t)$ as shown in Fig.~\ref{fig:diffusion} by LPSVD. 
The slopes of linear fits in Fig.~\ref{fig:diffusion3} (b) reproduce the experimental observation that the $f_i$ softening for XRD is larger than ARPES.
We use this to calculate the quantity $\frac{(d\tilde{f}_{\textrm{SS}}/dF)}{(d\tilde{f}_{\textrm{XRD}}/dF)}$, the second term of Eq.~\ref{eq:first two factors} right hand side.
It might seem counter-intuitive that for a given fluence, the phonon softening is larger in an XRD measurement while the phonon amplitude is larger in an ARPES measurement. This is because phonon amplitudes are quantified by their instantaneous behavior near $t=0$. Diffussion effects are limited to the the timescale of phonon generation of approximately one phonon period. In contrast, the analysis of phonon frequencies requires multiple phonon cycles. This allows for diffusion to act longer, which leads to the larger carrier density in a thin film sample as compared to the surface area of a bulk.

To clearly show the correction factor $[\frac{(d\tilde{u}_{\textrm{SS}}/d\tilde{f}_{\textrm{SS}})}{(d\tilde{u}_{\textrm{XRD}}/d\tilde{f}_{\textrm{XRD}})}]^{-1}$,  we plot the simulated, normalized A$_{1g}$ amplitudes as a function of A$_{1g}^{(2)}$ redshift in Fig.~\ref{fig:diffusion3} (c). The slopes are directly used to calculate the quantity $\frac{(d\tilde{u}_{\textrm{SS}}/d\tilde{f}_{\textrm{SS}})}{(d\tilde{u}_{\textrm{XRD}}/d\tilde{f}_{\textrm{XRD}})}$. 
The slope of the ARPES measurement $(d\tilde{u}_{\textrm{SS}}/d\tilde{f}_{\textrm{SS}})$ is a factor of 2.8 (3.0) larger than the XRD measurement $(d\tilde{u}_{\textrm{XRD}}/d\tilde{f}_{\textrm{XRD}})$ for the  A$_{1g}^{(1)}$ (A$_{1g}^{(2)}$) mode.

\begin{figure} 
		\includegraphics{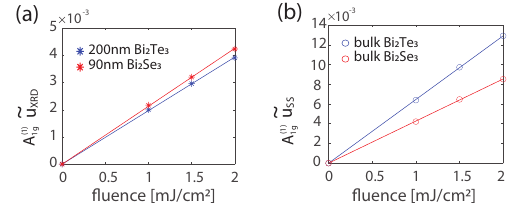}
		\caption{ (a) Simulated time-resolved XRD signal of the A$_{1g}^{(1)}$ response for Bi$_2$Te$_3$ film (200~nm) and Bi$_2$Se$_3$ film (90~nm), which correspond to the data taken at SACLA.
		(b) Simulated time-resolved ARPES signal of the A$_{1g}^{(1)}$ response for Bi$_2$Te$_3$ bulk and Bi$_2$Se$_3$ bulk. 
		}
		\label{fig:diffusion2}
	\end{figure}

Next we consider the carrier diffusion simulation for the DP calculation of Bi$_2$Se$_3$. Here, different samples are measured using the same experimental conditions and we can ignore the systematic errors induced by fluence calibration. In these simulations the absolute values of $\tilde{u}$ are irrelevant, only the ratios between them matter. Fig.~\ref{fig:diffusion2} shows that given the same absorbed fluence, the response of Bi$_2$Te$_3$ and Bi$_2$Se$_3$ differs, both for XRD in Fig.~\ref{fig:diffusion2} (a) and for ARPES in Fig.~\ref{fig:diffusion2} (b). Again, this is due to the difference in ambipolar diffusivities of  Bi$_2$Te$_3$ and Bi$_2$Se$_3$. 

\section{Ellispsometry}

	\begin{figure*} 
		\includegraphics{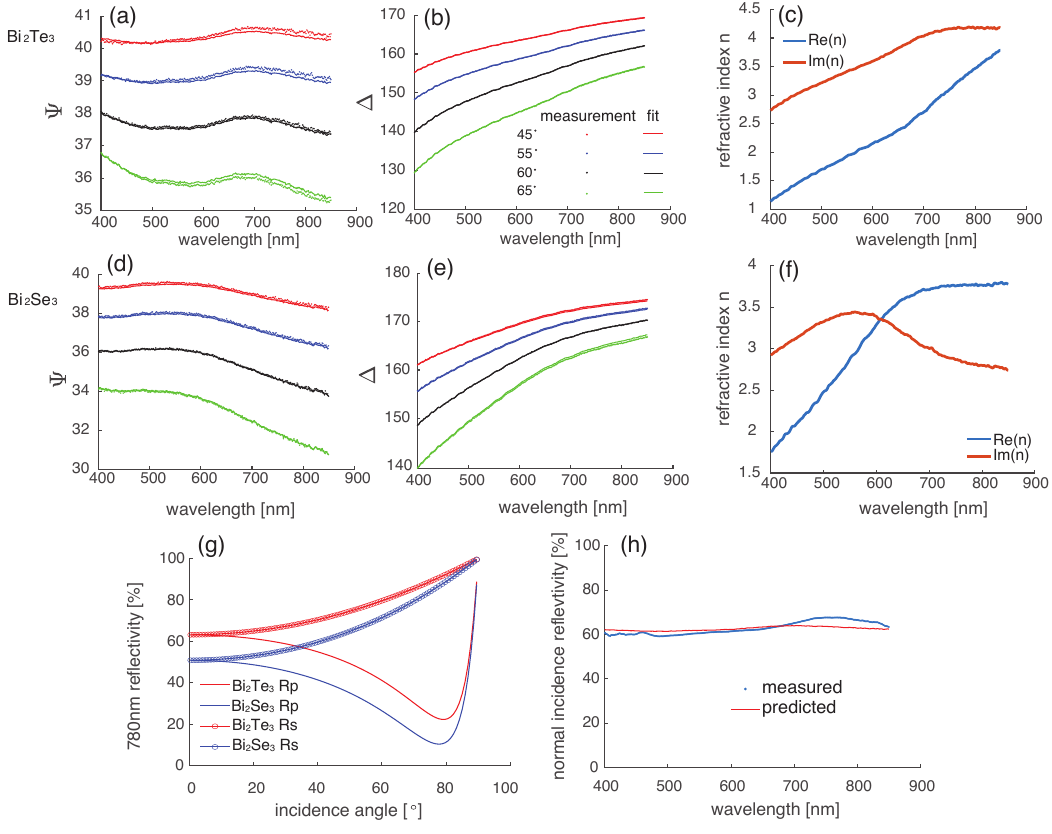}
		\caption{ Ellipsometry data for Bi$_2$Te$_3$ and Bi$_2$Se$_3$. (a) and (b) show  $\Psi$ and $\Delta$ from measurement (dots) and fit result (solid curve) for Bi$_2$Te$_3$. The reflectivity measurement was conducted at four different incidence angles. The fit uses a multilayer reflection model and takes into account the thickness of film and the refractive index of the substrate. The complex refractive index are shown in (c). 
		(d) (e) (f) Bi$_2$Se$_3$ ellipsometry results. 
		(g) The $s$ and $p$ reflectivity of 780~nm light as a function of incidence angle. 
		(h) shows the reflectivity of a 250~nm thick Bi$_2$Te$_3$ measured at normal incidence and the predicted reflectivity. 
		}
		\label{fig:ellipsometry}
	\end{figure*}

We measured ellipsometry data to obtain the complex refractive index  and in turn calculate the absorbed fluence in the Bi$_2$Te$_3$ and Bi$_2$Se$_3$ samples. 
Ellipsometry measures reflectivity with phase of $p$ and $s$ polarized light. The ratio of them, a complex number, is expressed as $\frac{r_p}{r_s}=\tan \Psi e^{i\Delta}$. In Fig.~\ref{fig:ellipsometry} we plot $\Psi$ and $\Delta$ as measured (dots) and as fitted (solid curve). 
Ellipsometry was conducted at four different incidence angles. The fit uses a multilayer reflection model and takes into account the thickness of the film as well as the refractive index of the substrate. 
The complex refractive indices as fit results are shown in Fig.~\ref{fig:ellipsometry} (e) and (f). 
To compare the pump energy density in different measurements using different incidence angles (XRD: 89.5$^{\circ}$, ARPES:50$^{\circ}$, optical: $<$5$^{\circ}$, grazing angle is defined as the angle between the beam and the sample normal), we calculate the reflectivity as a function of incidence angle in Fig.~\ref{fig:ellipsometry}(g), at a center wavelength of the pump laser of 780nm. 
In Fig.~\ref{fig:ellipsometry}(h) we show the reflectivity of a 250~nm thick Bi$_2$Te$_3$ measured at normal incidence and the reflectivity predicted from the measured refractive indices, showing good agreement.

\section{DFT methods }

\begin{figure} 
		\includegraphics{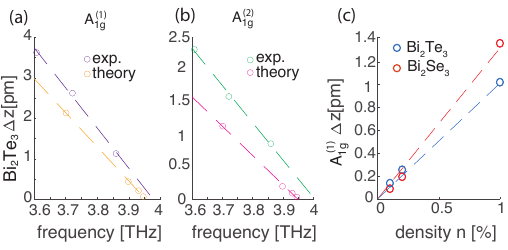}
		\caption{ (a) The constrained DFT simulations of photoexcited bulk Bi$_2$Te$_3$ atomic displacements for A$_{1g}^{(1)}$ as a function of the A$_{1g}^{(2)}$ frequency. 
		(b) Same as (a) but for A$_{1g}^{(2)}$ mode atomic displacements. 
		(b) cDFT simulation shows that Bi$_2$Te$_3$ and Bi$_2$Se$_3$ are driven at similar amplitudes given the same excitation density.
		}
		\label{fig:predict_amplitude}
\end{figure}

DFT calculations are performed in the local-density approximation (LDA)\cite{lda} employing the Hartwigsen-Goedecker-Hutter (HGH)
norm-conserving pseudopotentials \cite{hgh} using the ABINIT code\citep{abinit2009, abinit2016}. Spin-orbit coupling is included in all calculations. For electronic band structure calculations, Brillouin zone integrations are performed on a 12 $\times$ 12 $\times$ 1 Monkhorst-Pack $\textbf{k}$-points mesh in the slab calculations and 8$\times$8$\times$8 mesh in the bulk calculations. 
An energy cutoff for the plane waves of 15 Ha is used. 
For bulk phonon calculations  using density functional perturbation theory, a 12$\times$12$\times$12  and 6$\times$6$\times$6 $\textbf{k}$-points and $\textbf{q}$-points Monkhorst-Pack meshes are used to sample electronic and vibrational states, respectively. We use bulk phonons in all our calculations since they accurately represent the zone center A$_{1g}$ phonons in a slab with many quintuple layers.

In the bulk DFT calculations, we use the experimental lattice parameters given in Table \ref{table1}. In the slab calculations, we fully relax the atomic positions along the [111] direction, i.e. direction perpendicular to the surface, while keeping the in-plane lattice parameters at experimental values. 
We used 6-QL and 5-QL slabs for Bi$_2$Te$_3$ and Bi$_2$Se$_3$, respectively, to make sure that the energy gap between the surface states caused by the interaction between the two surfaces is smaller than a meV.

\begin{table}[h]
\begin{tabular}{|c|c|c|c|c|}
\hline
             & a & c & $\delta_1$ & $\delta_2$ \\ \hline
Bi$_2$Te$_3$ & 4.386 \AA~  & 30.497 \AA~  &  0.2095     &  0.4000     \\ \hline
Bi$_2$Se$_3$ & 4.143 \AA~  & 28.636 \AA~  & 0.2117    & 0.4008        \\ \hline
\end{tabular}
\caption{Bi$_2$Te$_{3}$ and Bi$_2$Se$_{3}$ lattice parameters taken from         Ref.\cite{nakajima1963}. $a$ and $c$ are the hexagonal lattice          constants, and $u$ and $\nu$ are the internal parameters                describing the position of the atoms inside the unit cell.}
\label{table1}
\end{table}

We calculate the forces in photoexcited bulk Bi$_2$Te$_3$ and Bi$_2$Se$_3$ immediately after photoexcitation using constrained DFT (cDFT) ~\cite{murray2005}, assuming that $n$ valence electrons are promoted to the conduction band. In cDFT calculation we assume one chemical potential, $i.e.$ electron and hole populations thermalize rapidly according to Fermi-Dirac distribution. For the photo-excited system, with electron-hole carrier density $n$, we obtain atomic forces $F_i$ in the $z$ or [111] direction on each atom $i$ in the unit cell as
\begin{equation}
    F_i(n) = F_i^0(n) + \sum_{j=1}^5 [z_j - z_j^0] \Phi_{ij}(n).
\end{equation}
The interatomic force matrix $\Phi_{ij}(n)$ depends on the carrier density and gives the phonon frequencies $\omega_\lambda(n)$ in the photo-excited system. 
The atomic forces $F_i^0(n)$ are the forces obtained in the photo-excited system for the equilibrium atomic positions $z_j^0$ in the ground state. The equilibrium atomic positions of the photoexcited system depend on $n$ and can be found from

\begin{equation}
    z_i^{equil}(n) = z_i^0 - \sum_{j=1}^5  \Phi_{ij}^{-1}(n) F_j^0(n).
\end{equation}

Fig.~\ref{fig:predict_amplitude} shows key results of the cDFT simulations. The atoms upon photoexcitation shift towards coordinates of higher symmetry (i.e., fractional coordinates move towards integer multiples of 1/5), which is consistent with the experimental observation of atomic motion direction deduced from the diffraction intensity change. 
The simulations reproduce the atomic motion amplitude of A$_{1g}$ modes as a function of the A$_{1g}^{(2)}$ frequency shift.  
They also correctly predict that Bi$_2$Te$_3$ and Bi$_2$Se$_3$ are driven at similar amplitudes given the same excitation density, consistent with the experimental results in Fig.~\ref{fig:SACLA_amplitude} (b).

\bibliographystyle{unsrt}
\bibliography{Bi2Te3Bi2Se3}
\end{document}